\documentclass[aps,prd,twocolumn,superscriptaddress,showpacs]{revtex4}

\usepackage{epsfig}
\usepackage{amssymb}
\usepackage{graphicx}
\usepackage{dcolumn}

\newcommand{\beq}{\begin{equation}}
\newcommand{\eeq}{\end{equation}}
\newcommand{\ds}{\displaystyle}
\newcommand{\beqar}{\begin{eqnarray}}
\newcommand{\eeqar}{\end{eqnarray}}

\begin{document}


\title{Study of $\pi\pi$ correlations at LHC and RHIC energies in 
       $pp$ collisions within the quark-gluon string model}

\author{M.S.~Nilsson}
\affiliation{
Department of Physics, University of Oslo, PB 1048 Blindern,
N-0316 Oslo, Norway
\vspace*{1ex}}
\author{L.V.~Malinina}
\altaffiliation[Also at ]{Department of Physics, University of Oslo, 
PB 1048 Blindern, N-0316 Oslo, Norway
\vspace*{1ex}}
\affiliation{
Skobeltzyn Institute for Nuclear Physics, Moscow State University, 
RU-119899 Moscow, Russia
\vspace*{1ex}}
\author{J.~Bleibel}
\altaffiliation[Also at ]{Max-Planck-Institut f\"ur Metallforschung,
Heisenbergstr. 3, D-70569 Stuttgart, Germany
\vspace*{1ex}}
\affiliation{
Institut f\"ur Physik, WA 331, Johannes Gutenberg Universit\"at Mainz,
D-55099 Mainz, Germany
\vspace*{1ex}}
\author{L.V.~Bravina}
\affiliation{
Department of Physics, University of Oslo, PB 1048 Blindern,
N-0316 Oslo, Norway
\vspace*{1ex}}
\author{E.E.~Zabrodin}
\altaffiliation[Also at ]{Skobeltzyn Institute for Nuclear Physics,
Moscow State University, RU-119899 Moscow, Russia
\vspace*{1ex}}
\affiliation{
Department of Physics, University of Oslo, PB 1048 Blindern,
N-0316 Oslo, Norway
\vspace*{1ex}}

\date{\today}

\begin{abstract}

The Quark Gluon String Model (QGSM) reproduces well the global 
characteristics of the $pp$ collisions at RHIC and LHC, e.g., 
the pseudorapidity and transverse momenta distributions at different 
centralities. The main goal of this work is to employ the Monte Carlo 
QGSM for description of femtoscopic characteristics in $pp$ collisions 
at RHIC and LHC. The study is concentrated on the low multiplicity and 
multiplicity averaged events, where no collective effects are expected.
The different procedures for fitting the one-dimensional correlation 
functions of pions are studied and compared with the space-time 
distributions extracted directly from the model. Particularly, it is 
shown that the double Gaussian fit reveals the contributions coming 
separately from resonances and from directly produced particles. The 
comparison of model results with the experimental data favors decrease 
of particle formation time with rising collision energy.

\end{abstract}


\maketitle

\section{\label{sec:intro}Introduction}

Experiments at Relativistic Heavy Ion Collider (RHIC) have 
demonstrated that hot and dense matter with partonic collectivity has 
been formed in ultrarelativistic Au+Au collisions at $\sqrt{s} = 
200$\,{\it A}GeV \cite{RHIC_QGP}. Proton-proton collisions are 
conventionally used as a reference to compare with nuclear collisions 
and to understand the observed collective effects. The new interest 
in general features of $pp$ collisions at ultrarelativistic energies 
appeared after the first publications of Large Hadron Collider (LHC)
data obtained in $pp$ interactions at $\sqrt{s} = 900$\,GeV and 
7\,TeV \cite{LHC_900_7_A,LHC_900_7_C}. 

The Bose-Einstein (BE) enhancement in the production of two identical 
pions at low relative momenta was first observed in $\bar{p}p$ 
collisions about 50 years ago \cite{GGL}. Since then, the developed 
correlation method, colloquially known at present as ``femtoscopy 
technique'', was successfully applied to the measurement of space-time 
characteristics of the production process at the distances of few 
fermis ($1\, fm\, =\, 10^{-15}\,m$) (see, e.g., 
\cite{pod89,led04,lis05} and references therein). The space-time 
relative distances are ``measured'' by femtoscopy studies at the 
points where the particles stop to interact. This moment occurs at the 
very late stage of the collision, long after the quark-gluon plasma 
(QGP) or any other exotic state of matter has disappeared. But signals 
like the geometric growth of the reaction zone and the specific 
features of the collective flow, generated by QGP pressure gradients, 
are imprinted in the final state as very specific space-momentum 
correlations influencing particle spectra and femtoscopic radii. 

The system created in ultrarelativistic $pp$ collisions at RHIC and 
especially at LHC energies can be similar to the system created in 
non-central heavy-ion collisions because of the large energy deposited 
in the overlapping region and therefore can also demonstrate 
collective behavior. The strong argument supporting this point of 
view comes from the observation of the almost identical multiplicity
and momentum dependencies of the femtoscopic radii in $pp$ and Au+Au 
collisions by the STAR collaboration at RHIC \cite{CF_STAR}.
In particular, the momentum dependence of the radii can be linked to
the collective flow developed in the system \cite{lis05}. The
striking result obtained by the ALICE collaboration from study of
the BE correlations in $pp$ collisions at $\sqrt{s} = 900$\,GeV
\cite{CF_ALICE} is the absence of the transverse momentum dependence,
whereas the event multiplicity dependence holds.

The aim of the present article is to study hadronization processes 
in $pp$ collisions at ultrarelativistic energies using momentum 
correlation technique within the Monte Carlo quark-gluon string
model (QGSM) \cite{mc_qgsm,QGSM_LHC} and to compare results of
calculations with the experimental data obtained at RHIC and LHC. 
This model describes successfully the main characteristics of $pp$
interactions, such as multiplicity, transverse momentum and 
(pseudo)rapidity distributions in a broad energy range from 
$\sqrt{s} = 200$\,GeV up to top LHC energy $\sqrt{s} = 7$\,TeV
\cite{QGSM_LHC}. We try to understand to what extent one is able 
to describe the correlation functions (CFs) in ultrarelativistic 
$pp$ collisions within the pure string model picture.

The paper is organized as follows. A brief description of the model
features is presented in Sect.~\ref{model}. Special attention is given
to the concept of the formation time, which plays a very important 
role for study of the femtoscopy correlations. Section~\ref{CF_repr}
introduces the method of correlation functions employed by both the
STAR and the ALICE collaboration. Model results obtained for $pp$
collisions at $\sqrt{s} = 200$\,GeV and $\sqrt{s} = 900$\,GeV are
presented in Sect.~\ref{results}. Comparison with the available 
experimental data is given as well. The proper choice of the baseline
is discussed, and ability of the double Gaussian fit to restore the
contributions of string processes and resonances to the correlation
functions is demonstrated. Finally, conclusions are drawn
in Sect.~\ref{concl}.

\section{The model}
\label{model}

\subsection{Basic features}
\label{subsec2_1}

Our model is the Monte Carlo realization of the quark-gluon string 
model developed in \cite{QGSM}. Similarly to the dual parton model
(DPM) \cite{DPM}, QGSM is based on Gribov's Reggeon field theory (GRT) 
\cite{GRT} accomplished by a string phenomenology of particle 
production in inelastic hadron-hadron collisions. To describe 
hadron-nucleus and nucleus-nucleus collisions the cascade procedure 
of multiple secondary interactions of hadrons is implemented. The 
model incorporates string fragmentation, formation of resonances, 
and rescattering of hadrons. As independent degrees of freedom QGSM 
includes octet and decuplet baryons, octet and nonet vector and 
pseudoscalar mesons, and their antiparticles. The momenta and 
positions of nucleons inside the nuclei are generated in accordance 
with the Fermi momentum distribution and the Woods-Saxon density 
distribution, respectively.

Pauli blocking of occupied final states is taken into account.
Strings in the QGSM can be produced as a result of the color
exchange mechanism or, like in diffractive scattering, due to
momentum transfer. The Pomeron, which is a pole with an intercept 
$\alpha_P(0) > 1$ in the GRT, corresponds to the cylinder-type 
diagrams. The $s$-channel discontinuities of the diagrams, 
representing the exchange by $n$-Pomerons, are related to process of 
$2 k\, (k\leq n)$ string production. If the contributions of all 
$n$-Pomeron exchanges to the forward elastic scattering amplitude are 
known, the AGK cutting rules \cite{agk} enable one to determine the 
cross sections for $2k$-strings. The hard gluon-gluon scattering and 
semi-hard processes with quark and gluon interactions are also 
incorporated in the model via the so-called hard Pomeron exchange
\cite{QGSM_LHC,prd92}. The hard Pomeron is nowadays a standard 
feature attributed to a variety of GRT-based microscopic models,
such as DPM \cite{DPM}, PHOJET \cite{phojet} and EPOS \cite{epos}.
Its presence seems to be necessary to describe the rise of
multiplicity at midrapidity and $p_T$ spectra of secondaries in
$pp$ interactions at LHC energies within the QGSM \cite{QGSM_LHC}.
Further details of the MC version of QGSM and its extension to $A+A$ 
collisions can be found in \cite{mc_qgsm,QGSM_LHC,QGSM_flow}.

\begin{figure}[htb]
 \resizebox{\linewidth}{!}{
\includegraphics{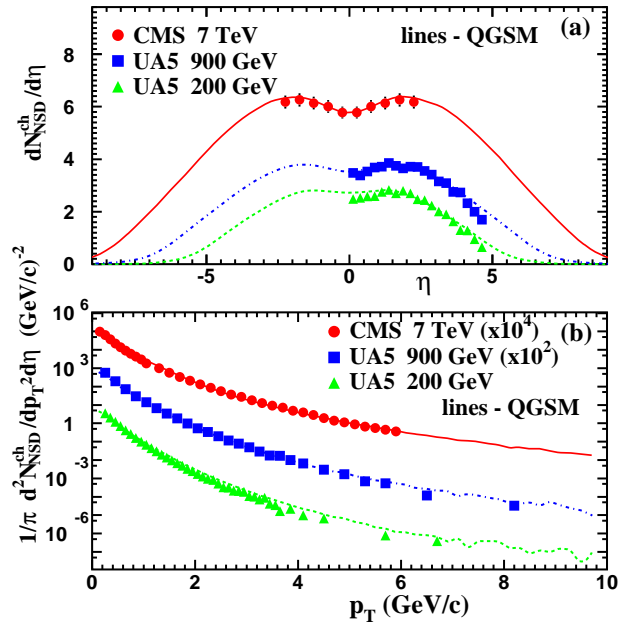}
}
\caption{
(Color online)
(a) The charged particle pseudorapidity spectra and (b) their
transverse momentum spectra in non-single-diffractive events
calculated in QGSM for $pp$ collisions at $\sqrt{s}= 200$\,GeV
(dashed lines), 900\,GeV (dash-dotted lines) and 7\,TeV (solid lines).
Symbols denote the experimental data taken from
\cite{UA5_rep,LHC_900_7_A,LHC_900_7_C}.
}
\label{fig:dndy_pt}
\end{figure}
Figure~\ref{fig:dndy_pt} displays the pseudorapidity and transverse
momentum distributions of charged particles produced in non-single
diffractive (NSD) $pp$ collisions at $\sqrt{s} = 200$\,GeV, 900\,GeV
and 7\,TeV, respectively. Experimental data are plotted onto the 
QGSM calculations also. Since the model reproduces the bulk 
characteristics of the collisions quite well, it would be useful to
apply the QGSM for the analysis of particle interferometry. Note,
however, that the GRT does not provide the space-time picture of
the system evolution, thus leaving room for the assumptions 
concerning the femtoscopy correlations quite open. Here one has to 
rely on approaches developed within the framework of the string
phenomenology. 

\subsection{QGSM and particle coordinates}
\label{subsec2_2}

The space-time evolution of the collisions starts from the 
interacting partons, i.e., quarks, diquarks and sea quarks distributed 
randomly in the projectile-target overlapping region. The strings 
between them are stretching and subsequently decaying into hadrons. 
Due to the uncertainty principle it takes time to create a hadron from 
constituent quarks. Also, hadrons are composite particles, and this
circumstance makes the definition of the formation time model
dependent. According to \cite{BG87}, two definitions of the formation
time or, equivalently, formation length are eligible in the framework
of the Lund string model \cite{lund}. In the ``yo-yo"
case it corresponds to the time/coordinate of the first intersection
point of the hadron constituents (``yo-yo'' formation time). In the
so-called constituent case it corresponds to the time/coordinate of
the point of rupture of the string (constituent formation time). In 
the present version of the QGSM the constituent formation time is 
used. The string length $L = M_s / 2\kappa$ depends on its mass $M_s$
and on the string tension $\kappa$. The mass of the string is not 
fixed. It is determined by the generation of longitudinal and 
transverse momenta of valence quarks at the string ends, that depend
on the momenta of colliding hadrons. The length of the string varies
from the maximum value determined by the momentum of the incident
hadron to the minimum value determined by the pion mass. Therefore,
for the formation of a resonance the mass and length of the string
must be much larger than for production of a pion.
 
The formation time $t_i^{*}$ and coordinate $z_i^{*}$ of $i$-th 
hadron in the string center of mass can be expressed via its energy 
$E_i^{*}$, its longitudinal momentum $p_{zi}^{*}$ and the 
longitudinal momenta/energies of all hadrons produced by the decay 
of this string
\begin{eqnarray} 
\label{eq:tLUND} 
t_{i}^{*} &=& \frac{1}{2 \kappa}\left( M_s-2\sum_{j=1}^{i-1}
p_{zj}^{*} \right) \ ,  \\
\label{eq:zLUND} 
z_{i}^{*} &=& \frac{1}{2 \kappa}\left( M_s-2\sum_{j=1}^{i-1}
E_{j}^{*} \right)\ .
\end{eqnarray} 
Then we calculate $t_i$ in the laboratory frame and make the 
propagation of the coordinates to this point $(x_i,y_i,z_i,t_i)$: 
$a_i=a_{0i}+t_ip_{ai}/E_i$, $a={x,y,z}$.
The initial spatial distribution of partons in a proton is found to
be insignificant for the pion coordinate distributions at freeze-out,
which are dominated by both the formation time of hadrons and decay
lengths of resonances. To study the possible reduction of the
formation time because of, e.g., increase of the string tension with
rising incident energy we introduce in 
Eqs.~(\ref{eq:tLUND})-(\ref{eq:zLUND}) the
scaling parameter $\alpha$, i.e., $\kappa = \alpha \kappa_0$, where
$\kappa_0 = 0.88$\,GeV/fm is the default value of the string tension
coefficient in the QGSM found from comparison with experimental data
at lower energies \cite{mc_qgsm}. 
\begin{figure}[htb]
 \resizebox{\linewidth}{!}{
\includegraphics{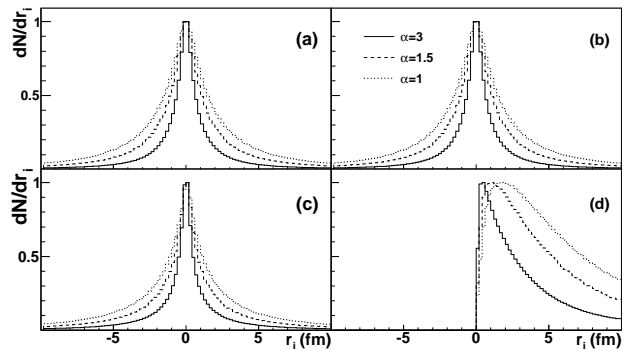}
}
\caption{
The $dN/dr_i,\ r_i = x\ {\rm (a)},\ y\ {\rm (b)},\ z\ {\rm (c)},\
t\ {\rm (d)}$ distributions of pions at freeze-out in $pp$ collisions
at $\sqrt{s}=900$\,GeV with $\alpha = 1$ (dotted line), $\alpha = 1.5$
(dashed line) and $\alpha = 3$ (solid line).
}
\label{fig:XYZTspatial900}
\end{figure}
The coordinate distributions of 
pions at freeze-out are shown in Fig.~\ref{fig:XYZTspatial900} for 
$pp$ collisions at $\sqrt{s}=900$\,GeV with $\alpha = 1,\ 1.5\ 
{\rm and}\ 3$. One can see that increase of $\alpha$ makes the 
coordinate distributions narrower.

\section{The correlation function representations}
\label{CF_repr}

The momentum correlations are usually studied with the help of
correlation functions of two or more particles. Particularly, the 
two-particle correlation function $CF(p_1,p_2)=A(p_1,p_2)/B(p_1,p_2)$ 
is defined as a ratio of the two-particle distribution from the same 
event $A(p_1,p_2)$ to the reference one. The reference distribution is 
typically constructed by mixing the particles from different events of 
a given class. 

In our simulations the weight of each particle pair is calculated 
according to quantum statistics (QS), using  particle four-momenta 
$p_i$ and four-coordinates $x_i$ of the emission points: $w = 1 + 
\cos(q \cdot \Delta x)$, where  $q = p_1 -p_2 $ and $\Delta x = x_1 - 
x_2$. The $CF$ is defined as a ratio of the weighted histogram of the 
pair kinematic variables to the unweighted one. The "ideal" case, 
$CF_{ideal}(p_1,p_2)=A(p_1,p_2, w)/A(p_1,p_2)$, uses unweighted pairs 
from the same events as the reference. In experiments one utilizes 
unweighted mixed pairs from different events as the reference, namely 
$CF_{realistic}(p_1,p_2)=A(p_1,p_2, w)/B(p_1,p_2)$. There is a 
difference between the ideal pair distribution $A(p_1,p_2)$ and the 
mixed one $B(p_1,p_2)$ due to presence of energy-momentum conservation 
for the pairs from the same event and absence of it in pairs from the 
mixed ones. This causes a smooth increase of $CF_{realistic}$ with $q$, 
which reflects the fact that due to energy-momentum conservation the 
probability of two particle emittence in the same direction is smaller 
than that in opposite directions. Therefore, a more complex fitting 
procedure is needed for the "realistic CF" than for the "ideal CF".

Generally, the correlations are measured as a function of pair 
relative momentum four vector $q$. An invariant form of this 
momentum difference commonly used in the one dimensional correlation 
analysis is $q_{inv} = \sqrt{q_{0}^{2} - |q|^{2}} $.
In both the STAR \cite{CF_STAR} and the ALICE \cite{CF_ALICE} 
experiments the correlation function is fitted to a single-Gaussian
\beq
CF_{single}(q_{inv}) = \left[ 1 + \lambda\exp{\left(
-R_{inv}^{2} q^{2}_{inv}\right)}\right]\,D(q_{inv}) \ ,
\label{eq:single}
\eeq
where the function $D(q_{inv})$ takes into account any non-femtoscopic 
correlations including the long-range correlations due to 
energy-momentum conservation described above. The parameters $R_{inv}$ 
and $\lambda$ describe the size of pion sources and the correlation 
strength, respectively. Here $R_{inv}$ is measured in the pair rest 
frame (PRF). Concerning the fit given by Eq.(\ref{eq:single}) we have 
to note that the best way to compare the model simulations with the 
experimental data is the direct comparison of the correlation functions.
Unfortunately, the CFs are not always available and one has to compare 
the results of the fit, that is more complicated. For instance, choice 
of the baseline $D(q_{inv})$ is rather arbitrary. The baseline should 
describe the CF behavior at large $q_{inv}$ where only the {\it
conservation laws\/} work, but the region of small $q_{inv}$ remains
terra incognita. Different experiments employ different extrapolations 
of the baseline to small $q_{inv}$, e.g., polynomial extrapolations, 
EMCIS-fit \cite{CF_STAR}, Monte Carlo simulations with PYTHIA and 
PHOJET \cite{CF_ALICE}, that give some specific behavior at small 
$q_{inv}$ due to strong jet contribution in these models, especially 
noticeable at large $k_t$. In order to reproduce the experimental 
fitting procedures in a model independent way and make a consistent 
comparison of our simulations with different experiments we will use 
below a flat baseline with $D(q_{inv}) = 1$ for STAR and ALICE data.

The correlation strength parameter $\lambda$ can differ from unity due 
to the contribution of long-lived resonances, particle 
misidentification and coherence effects. The 1D correlation functions 
were studied within the different ranges of the average pair 
transverse momentum $k_T = |\vec{p_{t,1}}+\vec{p_{t,2}}|/2$ in the 
mid-rapidity region. 

If large statistics sets are available it is possible to perform the
3D correlation analysis. Within realistic models, the directional and 
velocity dependence of the correlation function can be used to 
determine both the duration of the emission and the form of the 
emission region, as well as to reveal the details of the production 
dynamics \cite{pod89,led04,lis05}. For these purposes the correlation 
functions can be analyzed in terms of the $out$, $side$ and 
$longitudinal$ components of the relative momentum vector
${\bf q}=\{q_{out},q_{side},q_{long}\}$ \cite{pod83,bdh94}. Here 
$q_{out}$ and $q_{side}$ denote the transverse components of the 
vector ${\bf q}$, and the direction of $q_{out}$ is parallel to the 
transverse component of the pair three-momentum. The corresponding 
correlation widths are usually parametrized in terms of the Gaussian 
correlation radii $R_i$
\beqar \ds
 \label{eq:CF3D}
\nonumber
 && CF(p_{1},p_{2}) = \\
&&  1 + \lambda 
 \exp{\left( 
 - R_\mathrm{out}^2q_\mathrm{out}^2 
 - R_\mathrm{side}^2q_\mathrm{side}^2 
 - R_\mathrm{long}^2q_\mathrm{long}^2 \right)}\ .
\eeqar
The three dimensional analysis is performed in the longitudinal 
co-moving system (LCMS), where the pair momentum along the beam 
vanishes. It is possible to compare the radii measured in LCMS with 
$R_{inv}$ by making a boost of all radii from LCMS to PRF, namely,
$R_{out PRF}= \gamma_T R_{out}$, $R_{side PRF}= R_{side}$, 
$R_{long PRF}= R_{long}$ and averaging these radii.

The method used by STAR and ALICE experiments is to create a 3D 
correlation function by filling a three dimensional histogram with 
the full ${\bf q}=\{q_{out},q_{side},q_{long}\}$ vector in different 
ranges of the average pair transverse momentum 
$k_T = |\vec{p_{t,1}}+\vec{p_{t,2}}|/2$.

\section{Results and discussion}
\label{results}


\begin{figure}[htb]
 \resizebox{\linewidth}{!}{
\includegraphics{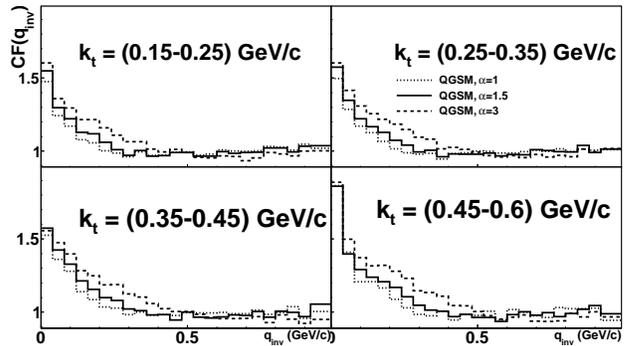}
}
\caption{
The $\pi^+ \pi^+$ CFs for $pp$ at $\sqrt{s}=200$\,GeV in four $k_T$
bins obtained by mixed pair reference distribution.
Cuts are $|\eta|<0.5$ and 0.12\,GeV/$c$ < $p_T$ < 0.8\,GeV/$c$, as in
the STAR experiment.
Calculations are performed with $\alpha = 1$ (dotted line), $\alpha =
1.5$ (solid line) and $\alpha = 3$ (dashed line).
}
\label{fig:CFsQGSMSTAR}
\end{figure}
\begin{figure}[htb]
 \resizebox{\linewidth}{!}{
\includegraphics{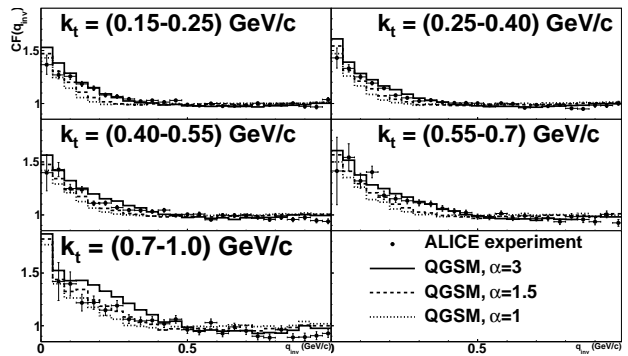}
}
\caption{
The same as Fig.\protect\ref{fig:CFsQGSMSTAR} but for $\sqrt{s} =
900$\,GeV/$c$. Cuts are $|\eta|<0.8$ and 0.15\,GeV/$c$ < $p_T$ <
1.0\,GeV/$c$. Calculations with $\alpha = 1$ (dotted line),
$\alpha = 1.5$ (dashed line) and $\alpha = 3$ (solid line) are
compared to ALICE results \protect\cite{CF_ALICE} with multiplicity
$M \le 6$.
}
\label{fig:CFsQGSMALICE}
\end{figure}

The two-pion correlation functions $CF(q_{inv})$ simulated for $pp$
collisions within the QGSM model with the scaling parameters $\alpha= 
1$, 2 and 3 are shown in different $k_T$ ranges in 
Fig.~\ref{fig:CFsQGSMSTAR} and Fig.~\ref{fig:CFsQGSMALICE} for 
$\sqrt{s} = 200$\,GeV and $\sqrt{s} = 900$\,GeV, respectively. The 
denominator of the CF was calculated by means of the mixing procedure 
described in Sec.~\ref{CF_repr}. As expected, smaller formation times
lead to smaller freeze-out radii of the particle sources and, 
therefore, to larger CFs in the interval $0 \leq q_{inv} \leq 
0.5$\,GeV/$c$. In Fig.~\ref{fig:CFsQGSMALICE} the correlation functions 
obtained with the QGSM are directly compared to those measured by the 
ALICE collaboration. The ALICE analysis performed for the minimum bias 
event sample gives for the value of the average pseudo-rapidity density 
$\langle dN_{ch}/d\eta \rangle = 3.6$, that coincides with the results 
of the QGSM simulations. We compare the QGSM low multiplicity sample 
with the ALICE data at low multiplicity bin $M \leq 6$. The best 
description is achieved for the scaling parameter equals to 3. In 
Fig.~\ref{fig:CFsQGSMALICE} one can see that the agreement between the 
\begin{figure}[htb]
 \resizebox{\linewidth}{!}{
\includegraphics{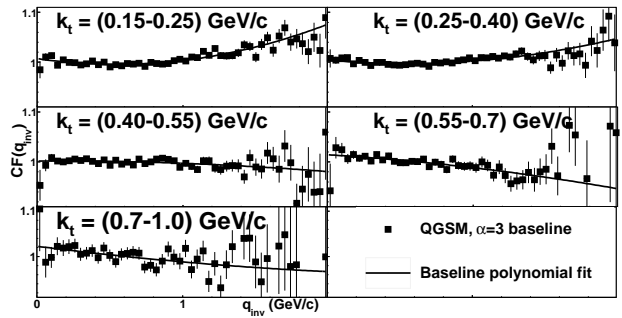}
}
\caption{
The baseline for $\pi^+ \pi^+$ correlation functions extracted from
QGSM calculations (full circles) of $pp$ collisions at $\sqrt{s} =
900$\,GeV in different $k_T$ intervals. Solid lines denote the fit
to polynomial $D(q_{inv})=a+bq_{inv}+c q_{inv}^{2}$.
}
\label{fig:baseline900}
\end{figure}
shapes of the correlation functions calculated within the QGSM and 
measured by the ALICE is rather good till $k_T<0.7$\,GeV/$c$. In the 
last $k_T$ bin $0.7 \leq k_T \leq 1.0$\,GeV/$c$ the experimental 
correlation function is about 15\% narrower than the QGSM one. To 
understand this effect better the realistic correlation functions 
without QS-weights, i.e.``baselines'', were constructed in different 
$k_T$ bins as displayed in Fig.\ref{fig:baseline900}. The 
energy-momentum conservation produces the long range correlation 
effects at large $q_{inv}$, for which the calculated values of the CFs
lie above the unity. In Ref.~\cite{CF_ALICE} the good description of
the long range correlations was obtained within the PYTHIA and PHOJET 
models. In Fig.\ref{fig:baseline900} the QGSM baseline $D(q_{inv})$  
demonstrates complicated behavior qualitatively similar to that of the
PYTHIA/PHOJET baselines but a bit flatter in low $q_{inv}$ interval 
for large $k_T$ bins.

\begin{figure}[htb]
 \resizebox{\linewidth}{!}{
\includegraphics{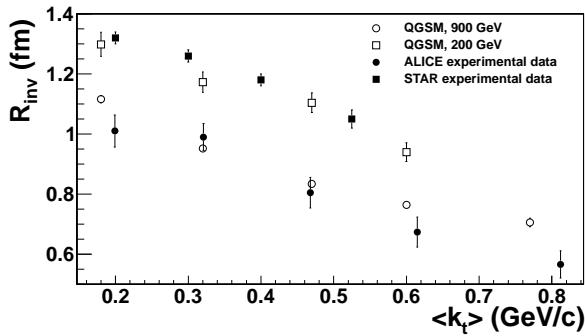}
}
\caption{
One-dimensional $\pi^+ \pi^+$ correlation radii as functions of $k_T$
in $pp$ collisions at $\sqrt{s} = 200$\,GeV (squares) and $\sqrt{s} =
900$\,GeV (circles). Open symbols denote STAR \protect\cite{CF_STAR}
and ALICE \protect\cite{CF_ALICE} experimental data, full symbols
present QGSM calculations with $\alpha = 1.5$ (200\,GeV) and
$\alpha = 3.0$ (900\,GeV), respectively. Both the model results and
the data are obtained from the fit to Eq.(\protect\ref{eq:single})
with the flat baseline.
}
\label{fig:RinvQGSM}
\end{figure}

Figure~\ref{fig:RinvQGSM} presents the $k_T$ dependence of $R_{inv}$ 
obtained from the fit to Eq.~(\ref{eq:single}) with the flat baseline 
of the QGSM CFs, shown in Fig.~\ref{fig:CFsQGSMSTAR} and  
Fig.~\ref{fig:CFsQGSMALICE}. The available STAR and ALICE 
data points with flat baseline \cite{CF_STAR,CF_ALICE} are 
averaged over the multiplicity and compared with the multiplicity 
averaged QGSM correlation functions. The best agreement with the STAR 
data \cite{CF_STAR} was obtained for calculations with $\alpha = 1.5$.

It was reported in \cite{CF_ALICE} that if PHOJET/PYTHIA baselines 
are chosen the correlation radii are practically independent on $k_T$ 
within the studied transverse momentum range, however, the strength of 
the $k_T$ dependence relies heavily on the baseline hypothesis. The
ALICE conclusion about the absence of $k_T$ dependence is based on
the assumption that both PHOJET and PYTHIA correctly describe the 
non-femtoscopic effects at low-$q_{inv}$ possibly related to minijets.
In this case the enhancement at low-$q_{inv}$ in the large $k_T$ bins 
is misinterpreted as Bose-Einstein enhancement. Since we assign 
Bose-Einstein weights to all pion pairs, we automatically imply that 
the enhancement at low-$q_{inv}$ is caused solely by the BE 
correlations. In such a case it will be improper to use the 
PHOJET/PYTHIA or our own QGSM baseline to exclude the assumed 
non-femtoscopy correlations at low $q_{inv}$. The rather successful 
description of the ALICE points within such approach suggests that 
there are no room for non-femtoscopic correlations at low $q_{inv}$ 
up to $k_T<0.7$\,GeV/$c$. 

The ALICE and STAR data points obtained with the flat baseline reveal 
a similar slope in Fig.~\ref{fig:RinvQGSM}, which is described rather 
well by the QGSM calculations with the scaling factors $\alpha = 1.5$ 
and $\alpha = 3$, respectively. However, the higher $k_T$-bins have 
larger deviations from the experimental points.

\begin{figure}[htb]
 \resizebox{\linewidth}{!}{
\includegraphics{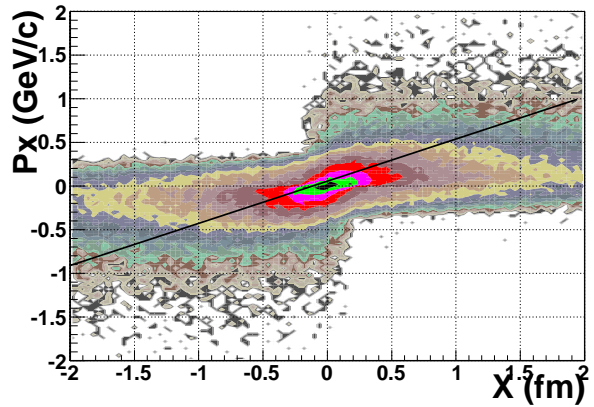}
}
\caption{(Color online)
The space-momentum correlations of direct pions produced in QGSM
calculated $pp$ collisions at $\sqrt{s} = 900$\,GeV.
Line is drawn to guide the eye.
}
\label{fig:XP}
\end{figure}

It is helpful to understand the origin of the strong $k_T$ dependence 
of the correlation radii in the QGSM model. The LUND hadronization 
schema described by Eqs.~(\ref{eq:tLUND})-(\ref{eq:zLUND}) introduces 
automatically the space-momentum correlations. The ``p-x'' 
correlations for the direct pions displayed in Fig.~\ref{fig:XP} 
look similar to the space-momentum correlations in hydrodynamic 
models, where they arise due to transverse collective flow. Note that 
only the particles with nearby velocities in their center-of-mass 
system contribute to the correlation function. If the ``p-x'' 
correlations are absent, the whole source is "seen" by the CF in any 
chosen $k_T$ range. Thus, there should be no $k_T$-dependence of the 
correlation radii.
In the presence of the ``p-x'' correlations the particles with close 
momenta come from nearby space regions of the source. Therefore, one
is measuring not the real geometrical size of the source, but rather
the size of the regions which emit particles of a given momenta, the 
so-called regions of homogeneity \cite{Sinyukov}. Higher $k_T$ pairs 
should have narrower coordinate distributions due to larger ``focusing 
effect''. It originates from the fact that particles with large 
momenta fly away from each other much quicker than particles with 
small momenta, so in order to be correlated they have to be very close 
in the coordinate space.
\begin{figure}[htb]
 \resizebox{\linewidth}{!}{
\includegraphics{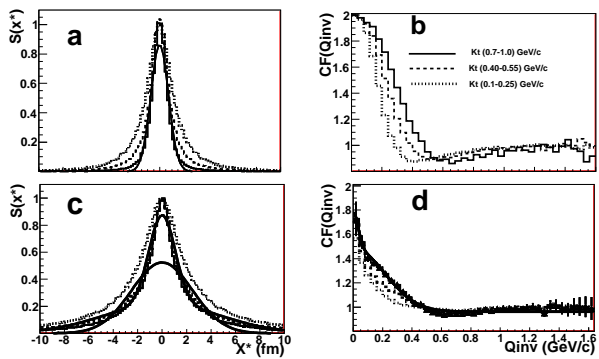}
}
\caption{
(a) Coordinate distributions of the direct pions in PRF in QGSM
calculated $pp$ collisions at $\sqrt{s} = 900$\,GeV with $\alpha = 3$.
The three transverse momentum intervals are $KT1= (0.1-0.25)$\,GeV/$c$
(dotted histogram), $KT3 = (0.4-0.55)$\,GeV/$c$ (dashed histogram) and
$KT5 = (0.7-1.0)$\,GeV/$c$ (solid histogram). The single Gaussian fit
for the KT5 bin is shown by the solid line.
(b) $CF_{ideal}$ for KT1 (dotted histogram), KT2 (dashed histogram)
and KT3 (solid histogram).
(c) The same as (a) but for all pions, the single and the double
Gaussian fits are shown for KT5 bin by the solid lines.
(d) The same as (b) but for all pions, double Gaussian fit is shown
for KT5 bin by the solid line.
}
\label{fig:XPRF}
\end{figure}
In Fig.~\ref{fig:XPRF}(a),(b) the transverse coordinate distributions 
are shown in the pair-rest-frame together with the corresponding 
correlation functions $CF_{ideal}$ for the direct pions in three 
$k_T$ ranges, namely $KT1 = (0.1-0.25)$\,GeV/$c$; $KT3 = 
(0.4-0.55)$\,GeV/$c$ and $KT5 = (0.7-1.0)$\,GeV/$c$. We see that
the widths of the $X_{PRF}$ distributions decrease with rising $k_T$ 
and the corresponding $CF_{ideal}$ become narrower. These widths are 
reproduced within error bars by the fit of $CF_{ideal}$ to the 
Gaussian given by Eq.~(\ref{eq:single}) with $D(q_{inv}) = 1$.

\begin{table}
\caption{
The fraction of pions from decay of main resonance species in QGSM 
and the path length $l^{*}$ of these states. 
}

\begin{tabular}{lcccc}\hline
          & $l^{*}$ (fm)    & 200~GeV              &900~GeV \\
\hline
Direct $\pi^{+}$                                  
          & -               &46.9\%                &37.5\%  \\
$\pi^{+}$ from $\rho^{0,+} \to \pi^{-,0}\pi^{+}$    
          & 3.3             &37.1\%                &40.7\%  \\
$\pi^{+}$ from $\omega \to \pi^{0}\pi^{-}\pi^{+}$   
          & 28.1            &11.2\%                &15.9\%  \\
$\pi^{+}$ from $K^{*,+} (\bar{K^{*,0}})\to K \pi^{+}$              
          & 8.0             & 4.2 \%               & 5.5 \% \\
\hline
\end{tabular}
\label{tab:ratio}
\end{table}

\begin{figure}[htb]
 \resizebox{\linewidth}{!}{
\includegraphics{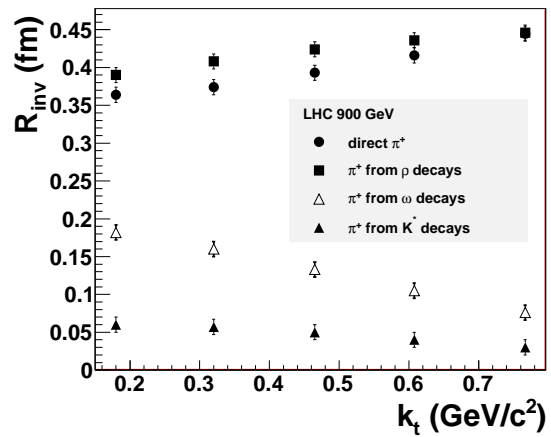}
}
\caption{
The evolution of the correlation radii $R_{inv}$ of identical pions
with their transverse mass $m_t$ in QGSM calculated $pp$ collisions
at $\sqrt{s} = 900$\,GeV. The symbols denote directly produced pions
(full circles) and pions coming from the decays of $\rho$ mesons
(full squares), $\omega$ mesons (open triangles) and $K^\ast$ (full
triangles), respectively.
}
\label{fig:resonances900}
\end{figure}

The important factor influencing the coordinate distributions is the 
ratio of direct pions to pions from resonance decay.
Table~\ref{tab:ratio} presents the fractions of pions from decay of 
the resonances most essentially contributed to the correlation
functions. The path length $l^{*} \simeq p_{d}/m_{\pi} \Gamma$
of these states in the c.m. frame of two identical pions at small 
value of $q_{inv}$ is listed in Table~\ref{tab:ratio} also. Here
$p_d$ is the momentum of the decay pion in the resonance rest frame 
\cite{LedProd}, $m_\pi$ is the pion mass and $\Gamma$ is the decay 
width. The pions from the decays of rather long-lived resonances 
$\omega$ and $K^*$ cause appearance of the exponential tails in the 
pion emission function, which distorts the Gaussian-like shape of the 
CF, see Fig.~\ref{fig:XPRF}(c),(d). Their relative contribution 
decreases with increasing $k_T$ due to kinematical reasons, whereas 
the relative contributions of direct pions and pions from $\rho$ 
decays increase as displayed in Fig.~\ref{fig:resonances900}. This 
effect also leads to decrease of the correlation radii with increasing 
$k_T$. The essentially non-Gaussian coordinate distributions that 
include contributions from resonances cannot be fitted well to a 
single Gaussian, however, the double-Gaussian fit reproduces its shape 
properly, see Fig.~\ref{fig:XPRF}(c). By fitting the corresponding 
CFs to a single Gaussian one cannot describe the narrow peak produced 
by pions from the resonance decays at low $q_{inv}$. On the other 
hand, using the double-Gaussian fitting procedure similar to the one 
suggested in \cite{LedProd}  
\beqar \ds
\nonumber 
&& CF_{double}(q_{inv}) = \left[ 1 + \lambda_1\exp{\left( 
 - R_{inv,1}^{2} q^{2}_{inv} \right)} \right. \\ 
&& + \left. \lambda_2 \exp{\left( -R_{inv,2}^{2} q^{2}_{inv}\right)} 
\right]\,D(q_{inv}) \ ,
\label{eq:double}
\eeqar
where parameters $R_{inv(1,2)}$ and $\lambda_{(1,2)}$ describe the 
sizes and the correlation strengths of the direct pion source and the 
one of the pions from the resonance decays, respectively, one gets 
much better description of the CF shape at low $q_{inv}$, as shown in 
Fig.~\ref{fig:XPRF}(d).

\begin{figure}[htb]
 \resizebox{\linewidth}{!}{
\includegraphics{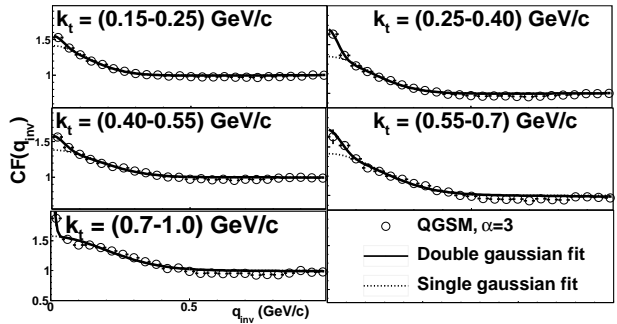}
}
\caption{
The fit of pion correlation functions, obtained in QGSM calculated
$pp$ collisions at $\sqrt{s} = 900$\,GeV with $\alpha = 3$, to
single Gaussian (dotted line) and double Gaussian (full line) in
five $k_T$ bins.
}
\label{fig:doubleG}
\end{figure}

In order to understand to what extent one is able to describe the 
correlation functions of all particles including the resonances by 
the different fitting procedures we make a comparison of the extracted 
values of $R_{inv}$ with the Gaussian widths of the coordinate 
distributions in the pair rest frame. The comparison is presented in 
Fig.~\ref{fig:XPRF}(c) for the ``ideal" correlation functions 
$CF_{ideal}$ and in Fig.~\ref{fig:doubleG} for the realistic CFs. 
The extracted parameters are listed in Table~\ref{tab:R1D} for three 
$k_T$ ranges, namely, $KT1 = (0.1-0.25)$\,GeV/$c$; 
$KT3 = (0.4-0.55)$\,GeV/$c$ and $KT5 = (0.7-1.0)$\,GeV/$c$.
Because of the sharp peak of the correlation functions at low
$q_{inv}$ the two radii restored by the double Gaussian fit vary
considerably. The first one is of the order of 1\,fm and has a
tendency to decrease with rising $k_T$, whereas the second one
is always larger than 3\,fm and increases to 13-14\,fm at high
transverse momenta. The second Gaussian is quite narrow thus
leading to a hair-width difference between the single-Gaussian
and double-Gaussian curves at $q_{inv} > 0.1$\,GeV/$c$.  

\begin{table}
\caption{
Comparison of the Gaussian widths of the coordinate distributions 
in PRF shown in Fig.\protect\ref{fig:XPRF}(c) $\sigma X_{PRF}^{all}$ 
for single and double Gaussian fit
with the $R_{inv}$ extracted by using different fitting strategies:
{\bf 1} - ``ideal'' CF is fitted to the single Gaussian 
Eq.(\protect\ref{eq:single}) with $D(q_{inv})=1$; \\
{\bf 2} - ``realistic'' CF is fitted to the single Gaussian 
Eq.(\protect\ref{eq:single}) with $D(q_{inv})=1$; \\
{\bf 3} -  ``realistic'' CF is fitted to the single Gaussian 
Eq.(\protect\ref{eq:single}) with $D(q_{inv})=a+bq_{inv} + 
c q_{inv}^{2}$; \\
{\bf 4} - ``realistic'' CF is fitted to the double Gaussian 
Eq.(\protect\ref{eq:double}) with $D(q_{inv})=1$; \\
{\bf 5} - ``realistic'' CF is fitted to the double Gaussian 
Eq.(\protect\ref{eq:double}) with $D(q_{inv})=a+bq_{inv} +
c q_{inv}^{2}$. \\ 
The selected transverse momentum intervals are
$0.1 \leq k_T \leq 0.25$\,GeV/$c$ (KT1),
$0.4 \leq k_T \leq 0.55$\,GeV/$c$ (KT3) and
$0.7 \leq k_T \leq 1.0 $\,GeV/$c$ (KT5), respectively.
}
\begin{tabular}{cccc}\hline
    Method    & \multicolumn{3}{c}{$R_{inv1(2)}$ (fm)}   \\
              &   KT1   &   KT3   &   KT5                \\
\hline
     1        &  1.00  &  0.77  &  0.66  \\ 
     2        &  1.26  &  0.84  &  0.71  \\
     3        &  1.10  &  0.84  &  0.71  \\
     4        &  1.23  &  0.81  &  0.71  \\
              &  5.04  &  3.26  & 13.97  \\
     5        &  1.05  &  0.81  &  0.71  \\
              &  3.61  &  3.25  & 13.83  \\
\hline
              &   &  $\sigma X_{PRF}^{all}$       & (fm)     \\
single G.     &  3.37  &  2.45  &  2.96  \\
double G.     &  1.48  &  1.08  &  1.00  \\
              &  5.35  &  4.72  &  4.23  \\
\hline
\end{tabular}
\label{tab:R1D}
\end{table}

\begin{figure}[htb]
 \resizebox{\linewidth}{!}{
\includegraphics{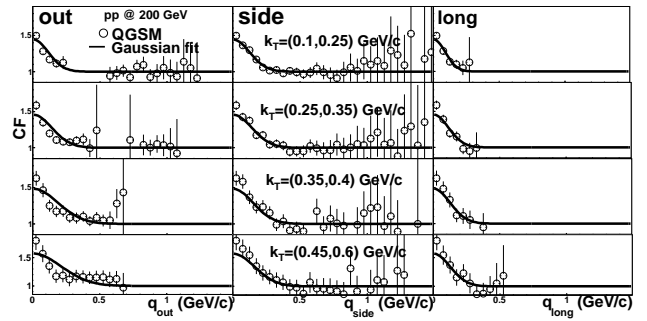}
}
\caption{
Projections of the 3D Cartesian representations of the correlation
functions onto the $q_{out}$, $q_{side}$, and $q_{long}$ axes, for
the minimum bias events from $pp$ collisions at 200 GeV for four
$k_T$ ranges. To project onto one $q$-component, the others are
integrated over the range $0 \leq q_i \leq 0.12$\,GeV/$c$.
}
\label{fig:CF3D_200GeV_full3D}
\end{figure}
\begin{figure}[htb]
 \resizebox{\linewidth}{!}{
\includegraphics{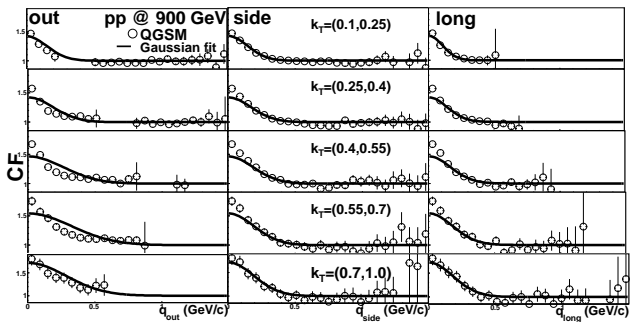}
}
\caption{
The same as Fig.\protect\ref{fig:CF3D_200GeV_full3D} but for the
low multiplicity bin $N_{\rm ch} < 11$ of $pp$ collisions at
900\,GeV in five $k_T$ ranges.
}
\label{fig:CF3D_900GeV_full3D}
\end{figure}

\begin{figure}[htb]
 \resizebox{\linewidth}{!}{
\includegraphics{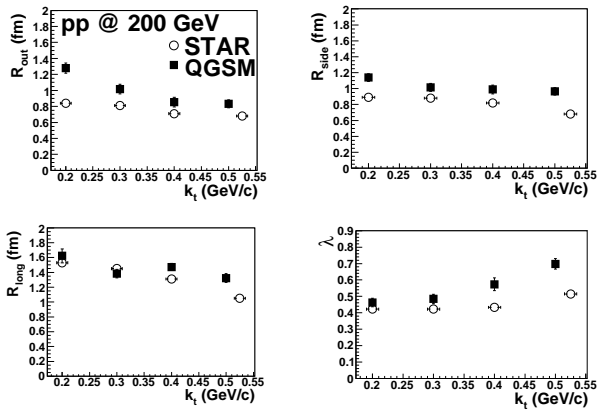}
}
\caption{
Three-dimensional $\pi^+ \pi^+$ correlation radii as functions of
$k_T$ in $pp$ collisions at $\sqrt{s} = 200$\,GeV for minimum bias
events. Open circles denote STAR experimental data, full squares
present QGSM calculations with $\alpha = 1.5$. Both the model
results and the data are obtained from the fit to
Eq.(\protect\ref{eq:CF3D}).
}
\label{fig:R3D_200GeV}
\end{figure}

\begin{figure}[htb]
 \resizebox{\linewidth}{!}{
\includegraphics{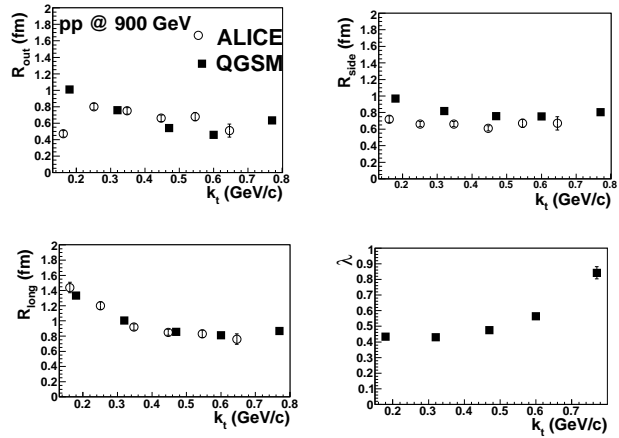}
}
\caption{
The same as Fig.\protect\ref{fig:R3D_200GeV} but for the low
multiplicity bin $N_{\rm ch} < 11$ of $pp$ collisions at $\sqrt{s} =
900$\,GeV. QGSM calculations are with $\alpha = 3$.
}
\label{fig:R3D_900GeV}
\end{figure}

The ideal 3D correlation functions for $\sqrt{s} = 200$\,GeV and 
$\sqrt{s} = 900$\,GeV, constructed for the minimum bias events and low 
multiplicity bin, are displayed in Fig.~\ref{fig:CF3D_200GeV_full3D} 
and Fig.~\ref{fig:CF3D_900GeV_full3D}, respectively. The calculations 
were done with $\alpha= 1.5$ and $\alpha= 3.0$, and the full 3D fit 
to the 3D Gaussian given by Eq.~(\ref{eq:CF3D}) was performed.
The extracted $R_{i}$ as functions of average $k_T$ are presented in 
Fig.~\ref{fig:R3D_200GeV} and Fig.\ref{fig:R3D_900GeV}. One can see 
that the experimental points are rather close to the QGSM ones 
especially for ALICE experimental data, see Fig.\ref{fig:R3D_900GeV}, 
where the low multiplicity bin is considered. Note that no integration 
over multiplicity was done in both cases.
At 200~GeV all radii demonstrate the weak decrease with $k_T$, whereas 
at 900~GeV the radii $R_{out}$ and $R_{side}$ are rather flat, the 
first point in $R_{out}$ is lower than the other ones, and only
$R_{long}$ demonstrates the decrease with rising $k_T$ as was observed 
by the ALICE collaboration at low multiplicity.

\section{Conclusions}
\label{concl}

The following conclusions can be drawn from our study. QGSM
calculations show strong dependence of the correlation radius on 
the transverse momentum of a pion pair. Similar dependence has been
observed by the STAR Collaboration, while the ALICE Collaboration
reported almost constant $R_{inv}$ with increasing $k_T$.
However, if the flat baseline is employed instead of the one
simulated by PYTHIA and PHOJET, the ALICE data demonstrate the
noticeable $k_T$ dependence as well. The origin of such a dependence 
in the QGSM is traced to the space-momentum correlations attributed 
to microscopic string models. If these correlations would be absent,
the correlation radius $R_{inv}$ will be independent on the pair
transverse momentum.

Pions in the model are produced either directly in the processes of
string fragmentation or from the decays of resonances. The relative
contribution of the long-lived resonances to pion emission function
decreases with rising $k_T$, while the corresponding contributions
of direct processes and short-lived resonances increase. Therefore,
the correlation radii of pions also decrease with increasing the 
pair transverse momentum. The fit of the 1D correlation functions 
to the double Gaussian provides a good description of the shape of 
the CFs at low-$q_{inv}$ range and enables us to separate the 
contributions from the direct pions and pions from the resonances.  

It was expected that the size of the freeze-out region in $pp$ 
collisions should increase with rising c.m. energy from $\sqrt{s} = 
200$\,GeV to $\sqrt{s} = 900$\,GeV due to the increase of interaction 
cross section and the number of produced resonances. Surprisingly, the 
radii measured by femtoscopy at $\sqrt{s} = 200$\,GeV are the same or 
even smaller than the ones at $\sqrt{s} = 900$\,GeV, as seen in
Fig.\protect\ref{fig:RinvQGSM}. The radii obtained within the standard 
LUND scenario of string breaking and the constituent formation time, 
implemented in the QGSM, appear to be larger compared to the 
experimental data. Our analysis favors reduction of the formation time 
with increasing energy of hadronic collision. One of the possible 
solutions is the process of string-string interaction via, e.g.,
fusion of strings that leads to increase of the string tension.

\begin{acknowledgments}
Fruitful discussions with A. Kisiel, R. Kolevatov, K. Mikhailov and
Yu. Sinyukov are gratefully acknowledged.
This work was supported by the Norwegian Research Council (NFR)
under contract No. 185664/V30.
\end{acknowledgments}

\end{document}